\documentclass[3p,times,procedia]{elsarticle}
\usepackage{nupha_ecrc}
\usepackage{graphicx}
\usepackage{float}
\usepackage[utf8]{inputenc}
\usepackage{bm}
\usepackage{hyperref}
\usepackage{amsmath}

\graphicspath{{./graphics/}}

\volume{00}

\firstpage{1}

\journalname{Nuclear Physics A}

\runauth{A.~Motornenko et al.}

\jid{nupha}

\jnltitlelogo{Nuclear Physics A}

\usepackage{amssymb}

\usepackage[figuresright]{rotating}

\begin{document}

\begin{frontmatter}


\dochead{XXVIIIth International Conference on Ultrarelativistic Nucleus-Nucleus Collisions\\ (Quark Matter 2019)}

\title{QCD equation of state at vanishing and high baryon density:\\
Chiral Mean Field model}


\author[ITP,FIAS]{Anton~Motornenko}
\author[FIAS]{Jan~Steinheimer}
\author[ITP,FIAS]{Volodymyr~Vovchenko}
\author[FIAS]{Stefan~Schramm$^\dagger$}
\author[ITP,FIAS,GSI]{Horst~Stoecker}
\address[ITP]{Institut f\"ur Theoretische Physik,
Goethe Universit\"at Frankfurt, D-60438 Frankfurt am Main, Germany}
\address[FIAS]{Frankfurt Institute for Advanced Studies, Giersch Science Center, D-60438 Frankfurt am Main, Germany}
\address[GSI]{GSI Helmholtzzentrum f\"ur Schwerionenforschung GmbH, D-64291 Darmstadt, Germany}
\let\thempfn\relax\footnote{$^\dagger$Deceased.}

\begin{abstract}
The thermodynamic properties of high temperature and high density QCD-matter are studied using the Chiral SU(3)-flavor parity-doublet Polyakov-loop quark-hadron mean-field model, CMF. The CMF model provides a proper description of lattice QCD data, heavy-ions physics, and static neutron stars. 
The behavior of lines of constant pressure with increase of baryon density is discussed. The rapid change of pressure behavior at $\mu_B/T\approx3$ suggests a strong contribution of baryons to thermodynamic properties at this region. The position of this region is very close to the radius of convergence for a Taylor expansion of the QCD pressure. The role of mesons and unstable hadrons in the hydrodynamic expansion of strongly interacting matter is also discussed.
\end{abstract}


\end{frontmatter}


\section{Introduction}
\label{sec:introduction}

The well established theory of strong interactions -- Quantum Chromodynamics (QCD) -- suffers calculation problems in the non-perutrbative regime: large coupling constant disfavors perturbative methods while the numerical sign problem prohibits lattice~(LQCD) calculations at finite densities. QCD phenomenology suggests a rather rich phase diagram of QCD matter, however no consensus on the structure of the phase diagram is reached yet. Albeit the hypothesis of a phase transition, attributed to the quark deconfinement, attracts a big interest in the scientific community. Definitive indications of a first-order phase transition are not yet found neither experimentally nor theoretically. The first principle LQCD calculations explore the QCD equation of state (EoS) in a restricted region close to $\mu_B=0$ where the baryon density $n_B$ vanishes. These detailed calculations find no signs of a sharp phase transition but a smooth crossover from hadronic to quark degrees of freedom. The extension of such calculations to finite baryonic densities are achievable by means of Taylor or fugacity series expansions in $\mu_B/T$, however those are limited by their radius of convergence. 

Here we present a mean-field approach to QCD thermodynamics applied for a wide range of densities and temperatures. We discuss the relevance of different degrees of freedom (baryons, mesons, hadron resonances, quarks) across the phase diagram predicted by the model.

\section{The CMF model}

\begin{figure}[h!]
\centering
\includegraphics[width=.45\textwidth]{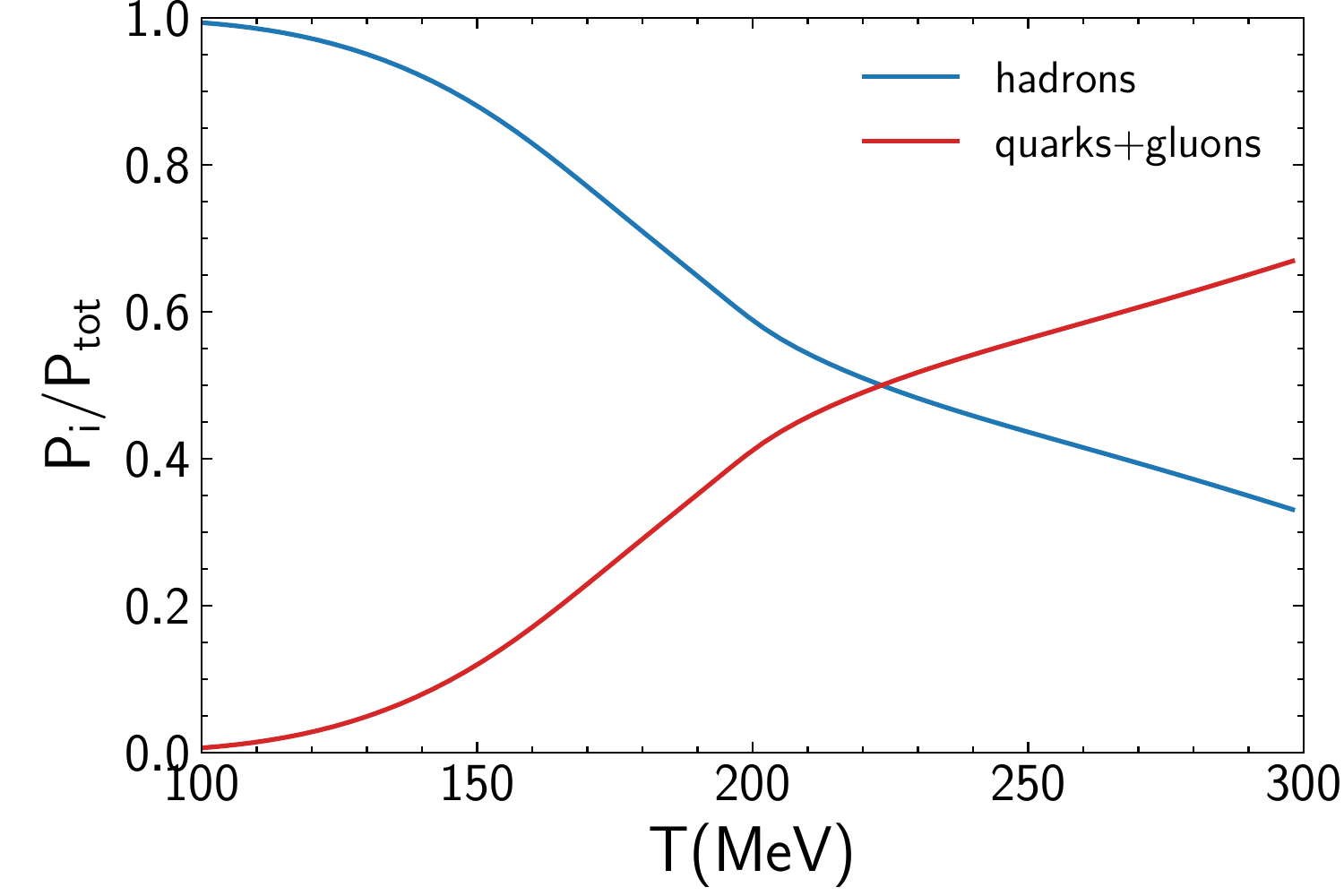}
\includegraphics[width=.45\textwidth]{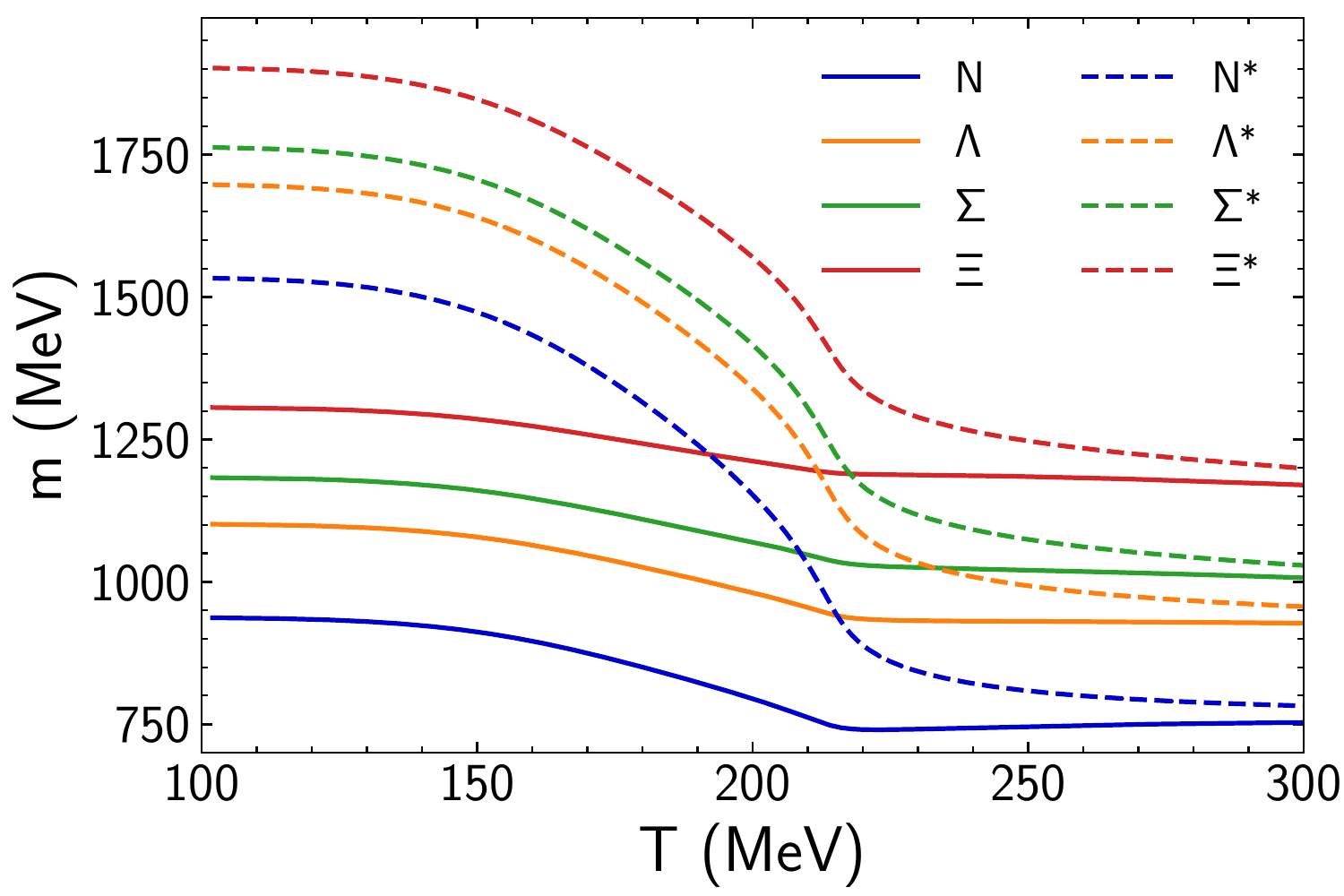}
\caption{{\bf Left:} contribution to the total pressure separately from quarks+gluons and from hadrons as functions of temperature $T$ at $\mu_B=0$. The suppression of quarks at low temperatures is provided due to the values of Polyakov loop parameter $\Phi$ and $\bar{\Phi}$, hadrons at high $T$ are suppressed by excluded volume interactions. {\bf Right:} masses of baryon octet and their parity partners as functions of $T$ at $\mu_B=0$, note smooth appearance of mass degeneracy at high $T$ between ground state baryons and their respective parity partners.}
\label{fig:mu0}  
\end{figure}

The Chiral SU(3)-flavor parity-doublet Polyakov-loop quark-hadron mean-field model, CMF \cite{Papazoglou:1998vr,Steinheimer:2010ib,Motornenko:2019arp} is a mean-field model for the statistical and thermodynamical properties of the QCD matter. The CMF model includes a complete list of known hadrons (with masses below 2.6 GeV) as well as the three light quark flavors. The transition between quark and hadron degrees of freedom as well as liquid vapor transition in nuclear matter are driven in the CMF by mean fields. The lowest baryon octet ($p, n, \Lambda, \Xi^-, \Xi^0, \Sigma^-, \Sigma^0, \Sigma^+$)  and their parity partners interact via mesonic mean fields (attractive scalar $\sigma,~\zeta$ and repulsive $\omega,~\rho,~\phi$ meson exchanges). The interaction with mesonic mean-fields ensure that the CMF model reproduces properties of nuclear matter \cite{Papazoglou:1998vr}. Parity doubling introduces heavy parity partners to the baryons of the lowest octet~\cite{Steinheimer:2011ea}. The effective masses of the parity partners depend on the chiral fields, therefore the partners become mass-degenerate as the chiral symmetry is restored:
\begin{eqnarray}
m^*_{i} = \sqrt{ \left[ (g^{(1)}_{\sigma i} \sigma + g^{(1)}_{\zeta i}  \zeta )^2 + (m_0+n_s m_s)^2 \right]} \pm g^{(2)}_{\sigma i} \sigma \pm g^{(2)}_{\zeta i} \zeta ~.
\label{eq:bar_mass}
\end{eqnarray}
This approach is supported by recent LQCD calculations, which show that the masses of the parity partners approach the same value above the pseudocritical temperature~\cite{Aarts:2018glk}.

The quarks are treated similar to the PNJL model for QCD matter~\cite{Fukushima:2003fw}. The effective quark mass $m_{q}^*$ is dynamically generated by the chiral fields $\sigma$ and $\zeta$ (non-strange and strange quark condensates). The quark contribution to the thermodynamic potential $\Omega_{q}$ is controlled by the Polyakov loop order parameter $\Phi$ and $\bar{\Phi}$:
\begin{equation}
\begin{aligned}
	\Omega_{q}=-VT \sum_{i\in Q}\frac{d_i}{(2 \pi)^3}\int{d^3k} \frac{1}{N_c}\ln\left(1+3\Phi e^{-\left(E_i^*-\mu^*_i\right)/T}+3\bar{\Phi}e^{-2\left(E_i^*-\mu^*_i\right)/T} +e^{-3\left(E_i^*-\mu^*_i\right)/T}\right),
	\label{eq:pnjl}
\end{aligned}
\end{equation}
where the masses of quarks are controlled by mesonic fields:
$m_{u,d}^*=-g_{q\sigma}\sigma+\delta m_q + m_{0q}\,, ~ m_{s}^*=-g_{s\zeta}\zeta+\delta m_s + m_{0q}$, the additional mass shift $\delta m_q$ for quarks is motivated by gluon contributions to the effective quark mass. It prevents quarks from appearing in the nuclear ground state. The values of $\Phi$ and $\bar{\Phi}$ are determined by the potential~\cite{Ratti:2005jh}:
	$U(\Phi,\bar{\Phi})=-\frac12 (a_0 T^4+a_1 T_0 T^3+a_2 T_0^2 T^2)\Phi\bar{\Phi}	+ b_3 T_0^4 \log[1-6\Phi\bar{\Phi}+4(\Phi^3 + \bar{\Phi}^{3})-3(\Phi\bar{\Phi})^2]$\,,
the parameters of the potential are tuned to the LQCD data so the CMF model reproduces LQCD thermodynamics at $\mu_B=0$~\cite{Motornenko:2019arp}.
Excluded volume corrections~\cite{Rischke:1991ke} ensure hadron suppression at high densities~\cite{Steinheimer:2010ib}:
\begin{eqnarray}
\rho_i=\frac{\rho^{\rm id}_i (T, \mu_i^*)}{1+\sum\limits_j v_j \, \rho^{\rm id}_j(T, \mu_j^*)} \, .
\end{eqnarray}
The excluded volume parameter $v_j$ is set to $v_B = 1$ fm$^3$ for all baryons and to $v_M = 1/8$ fm$^3$ for all mesons.

The phase diagram of the CMF model includes three critical regions, which are connected to the nuclear liquid-vapor phase transition, to the chiral symmetry restoration, and to the quark matter, respectively~\cite{Motornenko:2019arp}. The model predicts two first-order phase transitions. 
The first one is associated with the nuclear liquid-vapor phase transition and located at densities up to $n_0$. 
The second one appears at about four times the normal nuclear density due to the chiral symmetry restoration. The critical point temperatures for both transitions are similar: $T_{CP}\approx20$ MeV. The transition to quark matter takes place at higher densities and it is always a smooth crossover. 
The dynamics in the region of the phase diagram accessible to experiments of high energy heavy-ions collisions at the freeze-out stage is dominated by remnants of the nuclear liquid-vapor phase transition and resonances. 
Other critical regions predicted by the CMF may be probed by analysis of neutron star properties and binary neutron star mergers.

\section{The CMF phase diagram}
\begin{figure}[h!]
\centering
\includegraphics[width=.49\textwidth]{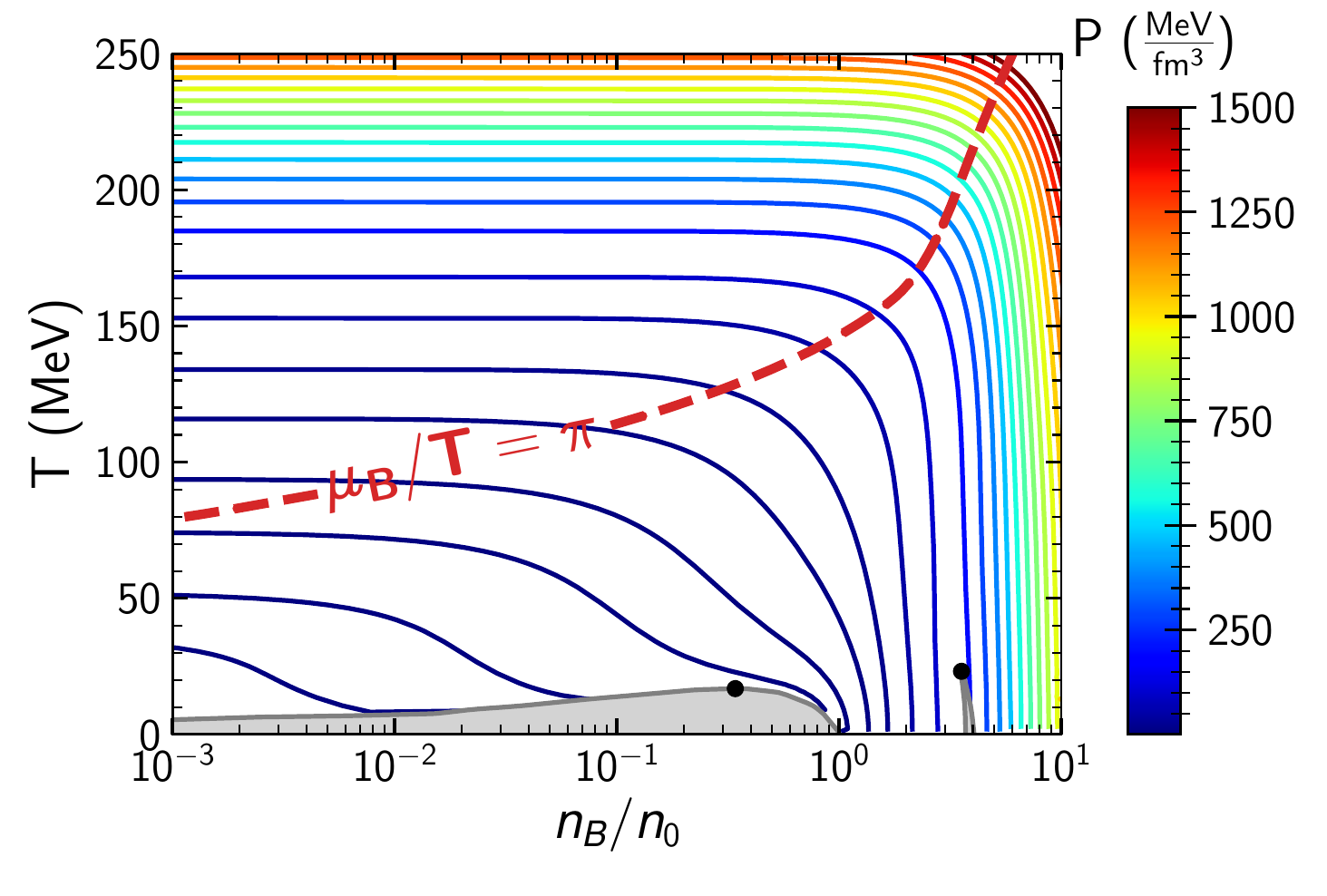}
\includegraphics[width=.49\textwidth]{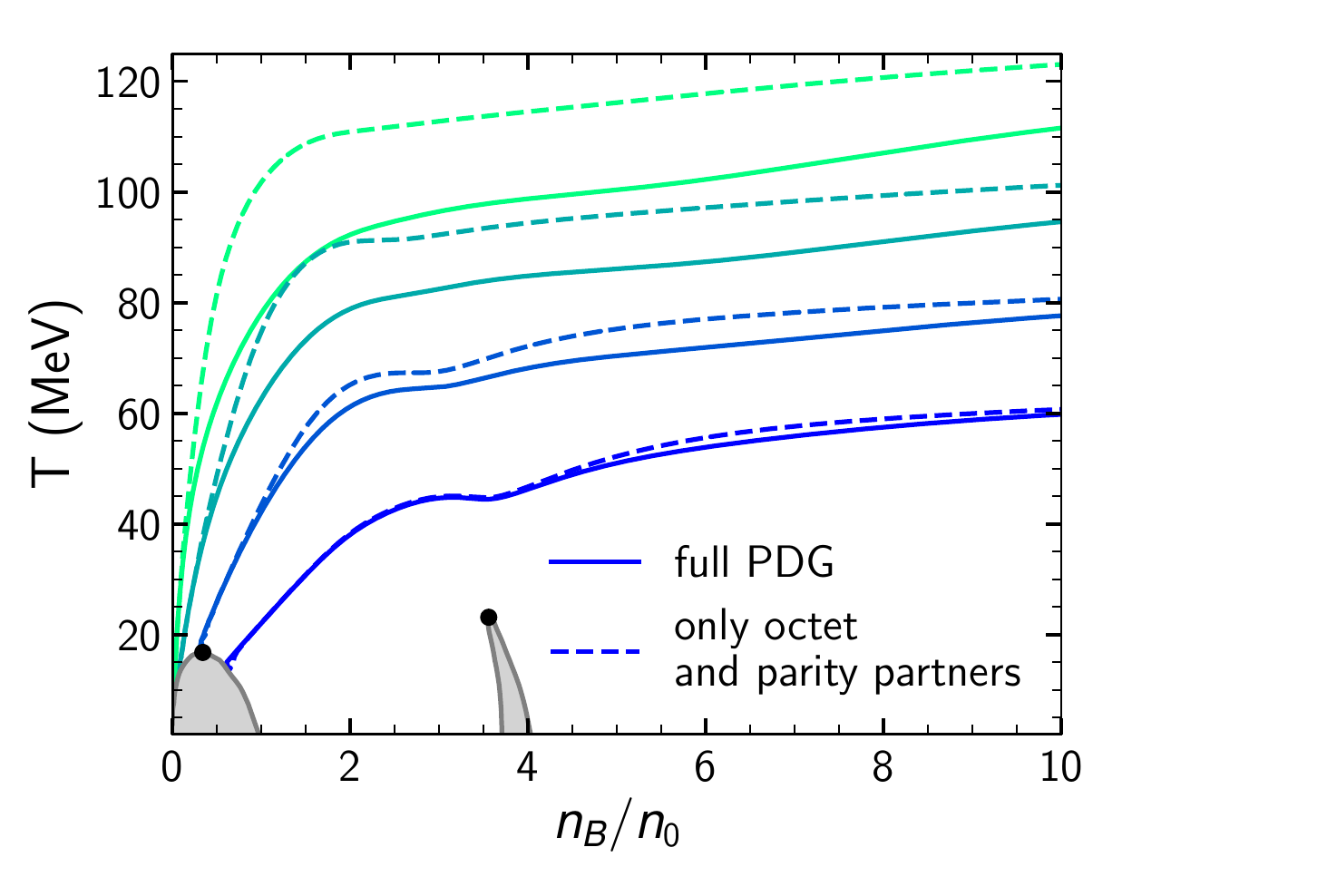}
\caption{{\bf Left:} Lines of constant pressure over $n_B-T$ plane, color indicates the value of pressure $P$ at the contour, baryon density $n_B$ is normalized to the nuclear saturation density $n_0$. The red dashed line indicates the line of $\mu_B/T=\pi$, this line lies in a proximity of abrupt tilt of constant pressure lines. The grey shaded regions illustrate mixed regions produced by nuclear liquid-vapor and chiral phase transitions. Black dots indicate critical endpoints. {\bf Right:} Isentropic trajectories, lines of constant entropy per baryon $S/A$, calculated within the CMF model with the full particle list (solid) and only with the stable baryons+quarks (dashed) where mesons and resonances are neglected. Note the increase of temperature for the isentropes where mesons and resonances are neglected.}
\label{fig:isobars}
\end{figure}

The LQCD studies of QCD matter phase diagram are mainly carried out via the Taylor expansion in series of $\mu_B/T$~\cite{Allton:2002zi}. 
The current estimates of a radius of convergence of the expansion suggest $R_{\mu_B/T} \sim 3$ at $T \gtrsim 135$~MeV~\cite{Bazavov:2017dus,Vovchenko:2017gkg}, which limits the applicability of the lattice studies of matter at large baryon densities. 
The CMF model allows to study physics also at much higher $\mu_B$.
On Fig.~\ref{fig:isobars} (left) the lines of constant pressure -- the isobars -- are presented across the $n_B-T$ plane. 
Additionally, the $\mu_B/T=\pi$ line is plotted, 
which is a convergence radius limitation that may arise due to the Roberge-Weiss~\cite{Roberge:1986mm} singularity at imaginary values of chemical potential, $\mu_B/T=i\pi$. 
$R_{\mu_B/T} \approx \pi$ values have been suggested by means of a cluster expansion model~(CEM) analysis of LQCD data~\cite{Vovchenko:2017gkg}. 
Fig.~\ref{fig:isobars} (left) illustrates that up to $\mu_B/T\approx\pi$ the pressure changes insignificantly with isobars being virtually horizontal. 
In this region the pressure is dominated by particles and antiparticles almost on the same level and for $T\gtrsim 100$ MeV is dominated by mesons which do not carry baryon charge. 
For temperatures $T>100$ MeV the convergence region significantly widens due to the increasing dominance of mesonic degrees of freedom which smears out the effects of finite $\mu_B$. 
At sufficiently large values of $T$ the matter is mostly composed by quarks and gluons with the latter not carrying the baryon charge. 
For quark-gluon matter the radius of convergence is expected to be limited by the Roberge-Weiss transition although the answer may depend on the exact nature of this transition~\cite{Lombardo:2005ks}. 
The presented behavior suggests that the Taylor expansion in $\mu_B/T$ in the region  $\mu_B\approx0$ only allows to explore matter where effects of finite baryon densities are small and pressure is dominated by mesons and gluons, the pressure of particles that carry baryon charge is on the same level as of antiparticles with a negative baryon charge. At temperatures $T\approx 150$ MeV the expansion is reliable up to $n_B\lesssim 2 n_0$ and decreases for smaller temperatures. This finite region of baryon densities restricts applications of the LQCD results  to low energy heavy ion collisions and neutron star mergers. 
Therefore, hydrodynamical simulations of such systems require the EoS produced by different means.

We also examine the role of mesonic and resonance degrees of freedom in hydrodynamic evolution of a hot, strongly interacting fireball. The relevance of mesons and hadronic resonances in the expansion of fireball created in heavy ion collisions was pointed out a long time ago~\cite{Stoecker:1978jr}. However, the role of these thermally excited states in hot hadronic matter created in neutron star mergers is still under investigation~\cite{Hanauske:2019qgs,Motornenko:2019lwh}.  Figure~\ref{fig:isobars} presents isentropes -- the lines of constant entropy per baryon $S/A=const$ -- which represent trajectories along $n_B$-$T$ plane of an ideal hydrodynamic expansion.
The trajectories show that an omission of mesons and baryonic resonances leads to an artificial increase of temperature in the hydrodynamic evolution.

\section{Summary}
We presented a general purpose model for the QCD thermodynamics: the CMF model. 
The model provides a mean-field level description of the $\mu_B=0$ LQCD data and properties of nuclear matter. 
It also predicts neutron star properties in agreement with current experimental constraints. 
We studied thermodynamics of the QCD matter at finite baryon densities and discussed it in the context of the  Taylor expansion around $\mu_B=0$ in powers of $\mu_B/T$. 
The calculations indicate that close to the $\mu_B/T=\pi$ value the behavior of isobars in $n_B-T$ plane is modified significantly. 
Assuming $\mu_B/T < \pi$ to be an approximate range of applicability of the Taylor expansion one can only reach densities $n_B<2n_0$ at temperatures $T<150$ MeV with this method. 
This small region of baryon densities thus limits the applicability of a LQCD computed EoS for hydrodynamic simulations of low energy heavy-ion collisions. 
We also discussed the role of mesonic and resonance degrees of freedom in hydrodynamic modeling of strongly interacting matter. 
If these states are omitted in an EoS used in simulations of neutron star mergers, this leads to an artificial increase of temperatures observed in such events.

The authors thank the BMBF, the DFG, HIC for FAIR, HGS-HIRe for FAIR, and the Samson AG for support. H.St. appreciates the support from J. M. Eisenberg Laureatus chair and the W. Greiner Gesellschaft.

\bibliographystyle{elsarticle-num}

\end{document}